\documentclass[sigconf,nonacm]{acmart}

\usepackage{kotex}
\usepackage{placeins}
\usepackage{float}
\usepackage{microtype}
\usepackage{enumitem}
\usepackage{booktabs}
\usepackage{array}
\usepackage{multirow}
\usepackage{tabularx}
\usepackage{subcaption}
\usepackage{balance}
\usepackage{hyperref}
\usepackage{pgfplots}
\usetikzlibrary{patterns,patterns.meta}
\pgfplotsset{compat=1.18}
\usepackage[normalem]{ulem}
\usepackage{pgfplotstable}
\usepackage{mathtools}
\useunder{\uline}{\ul}{}
\newcolumntype{C}{>{\centering\arraybackslash}p{2cm}}
\hypersetup{
    colorlinks=true,
    linkcolor=blue,
    urlcolor=cyan
}
\let\ul\undefined
\usepackage{soul}

\usepackage{tikz}
\usepackage{pgfplots}
\pgfplotsset{compat=1.18}
\usepackage{titlesec}
\usepackage{graphicx}
\usepackage{array}
\usepackage{diagbox}   
\usepackage{makecell}  
\usepackage{multirow}
\usepackage{booktabs}
\usepackage{adjustbox} 

\usepackage{balance}

\makeatletter

\titlespacing*{\section}
  {0pt}      
  {2.0ex}    
  {1.0ex}    
\makeatother

\definecolor{darkblue}{rgb}{0.0, 0.0, 0.55}
\definecolor{ao(english)}{rgb}{0.0, 0.5, 0.0}
\definecolor{coolblack}{rgb}{0.0, 0.18, 0.39}
\definecolor{purpleheart}{rgb}{0.41, 0.21, 0.61}
\definecolor{pastelviolet}{rgb}{0.8, 0.6, 0.79}
\definecolor{lightskyblue}{rgb}{0.53, 0.81, 0.98}
\definecolor{palecornflowerblue}{rgb}{0.67, 0.8, 0.94}
\definecolor{lightmauve}{rgb}{0.86, 0.82, 1.0}
\definecolor{lightpastelpurple}{rgb}{0.69, 0.61, 0.85}
\definecolor{bronze}{rgb}{0.8, 0.5, 0.2}
\definecolor{armygreen}{rgb}{0.29, 0.33, 0.13}
\definecolor{darkpowderblue}{rgb}{0.0, 0.2, 0.6}
\definecolor{falured}{rgb}{0.5, 0.09, 0.09}
\definecolor{outerspace}{rgb}{0.25, 0.29, 0.3}
\definecolor{tangerine}{rgb}{0.95, 0.52, 0.0}
\definecolor{seagreen}{rgb}{0.18, 0.55, 0.34}
\definecolor{springgreen}{rgb}{0.0, 1.0, 0.5}
\definecolor{applegreen}{rgb}{0.55,0.71,0.0}
\definecolor{amethyst}{rgb}{0.6,0.4,0.8}
\definecolor{amber}{rgb}{1.0,0.49,0.0}
\definecolor{darkgreen}{rgb}{0,0.4,0} 

\usepackage[table]{xcolor}
\usepackage{pgfplots}
\usetikzlibrary{plotmarks}
\usepgfplotslibrary{groupplots}

\AtBeginDocument{%
  }

\begin{document}

\title{E-MMKGR: A Unified Multimodal Knowledge Graph Framework for E-commerce Applications}

\author{Jiwoo Kang}
\affiliation{
	\institution{UNIST}
	\city{Ulsan}
  	\country{Korea}
}
\email{jiwoo0212@unist.ac.kr}

\author{Yeon-Chang Lee}
\authornote{Corresponding author.}
\affiliation{
	\institution{UNIST}
	\city{Ulsan}
  	\country{Korea}
}
\email{yeonchang@unist.ac.kr}

\begin{abstract}
Multimodal recommender systems (MMRSs) enhance collaborative filtering by leveraging item-side modalities, but their reliance on a fixed set of modalities and task-specific objectives limits both modality extensibility and task generalization. 
We propose \textbf{\emmkgr}, a framework that constructs an e-commerce–specific Multimodal Knowledge Graph (\emmkg) and learns unified item representations through GNN-based propagation and KG-oriented optimization. 
These representations provide a shared semantic foundation applicable to diverse tasks. 
Experiments on real-world Amazon datasets show improvements of up to 10.18\% in Recall@10 for recommendation and up to 21.72\% over vector-based retrieval for product search, demonstrating the effectiveness and extensibility of our approach.
The codebase of \emmkgr\ is available at \url{https://github.com/jiwoo0212/E-MMKGR_}.
\end{abstract}

\newcommand{\spec}{{\it spec.}}
\newcommand{\aka}{{\it a.k.a.}}
\newcommand{\ie}{{\it i.e.}}
\newcommand{\eg}{{\it e.g.}}

\newcommand{\emmkgr}{\textsc{\textsf{E-MMKGR}}}
\newcommand{\emmkg}{\textsc{\textsf{E-MMKG}}} 

\newcommand{\blue}{\textcolor{blue}}

\newcommand{\mj}[1]{\textcolor{blue}{[MJ: #1]}}
\newcommand{\jw}[1]{\textcolor{green}{[JW: #1]}}
\newcommand{\yc}[1]{\textcolor{red}{[YC: #1]}}

\maketitle
\section{Introduction} \label{sec:intro} \noindent\textbf{Background.}
With the rapid growth of e-commerce, recommender systems play a crucial role in product discovery. 
Traditional collaborative filtering (CF), however, relies solely on user–item interactions, 
making it vulnerable to sparsity and cold-start issues.
To alleviate these limitations, \textbf{multimodal recommender systems (MMRSs)}~\cite{grcn,dualgnn,lattice,bm3,freedom,lgmrec, mmrec_survey} have incorporated product-side modalities such as images and text to enrich item representations.

\vspace{1mm}
\noindent\textbf{Limitations.}
Despite recent progress, current MMRSs still struggle to handle
the growing diversity of modalities and user intent.

\textbf{(L1) Limited Modality Extensibility.}
New forms of product information continue to appear in e-commerce.
Beyond images and descriptions, data such as user reviews or LLM-generated content (\eg, image captions or product summaries) now play an important role.
However, most existing MMRSs~\cite{grcn,dualgnn,lattice,bm3,freedom,lgmrec, mmrec_survey} are designed around a fixed set of modalities, and adding new ones usually requires nontrivial changes to the overall architecture, limiting their ability to adapt to the evolving modality landscape.

\begin{figure}[t]
\vspace{0.5cm}
\centering
\includegraphics[width=\linewidth]{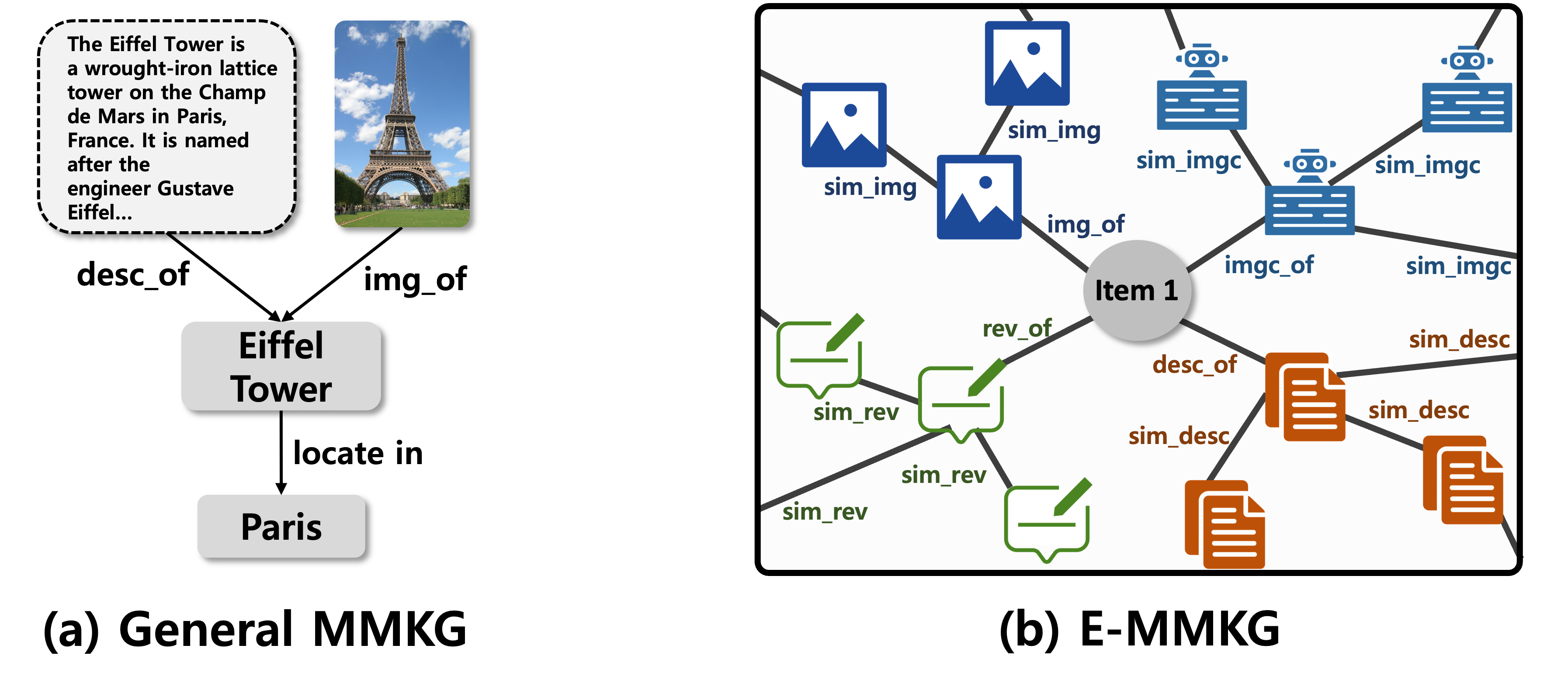}
\caption{General MMKG and the proposed E-MMKG.}
\label{fig:intro}
\end{figure}

\textbf{(L2) Limited Task Generalization.}
User expectations in e-commerce have expanded from receiving a simple ranked list to exploring and comparing products across multiple dimensions.
Supporting such diverse user intents requires models to capture the underlying item semantics rather than optimizing for a single task.
However, existing MMRSs learn task-specific item representations that fail to generalize to related tasks, including similar-item discovery or query-based product search, or attribute-aware filtering, limiting their applicability beyond recommendation.

\vspace{1mm}
\noindent\textbf{Motivation.}
To address these limitations, it is natural to ask:
``\textit{What kind of item representation can unify diverse modalities and user intents within a single semantic space?}''
Our fundamental motivation for this question is that many e-commerce tasks, though they differ in their objectives, rely on a \textbf{shared need to ground multimodal item information in a unified, relational semantic space.}
Such a representation offers a shared foundation for recommendation and search and naturally extends to tasks beyond them, while also supporting modality extensibility within a unified framework.

\vspace{1mm}
\noindent\textbf{Our Work.}
Therefore, in this work, we aim to build a unified representation framework that supports diverse modalities and user intents in e-commerce. 
To this end, we propose \textbf{\emmkgr}, an \underline{E}-commerce–specific semantic framework that constructs a structured \underline{M}ulti\underline{M}odal \underline{K}nowledge \underline{G}raph and learns unified item \underline{R}epresentations applicable to recommendation, search, and beyond.

Our design is inspired by general \textit{Multimodal Knowledge Graphs} (MMKGs)~\cite{mmkgsurvey} (see Figure~\ref{fig:intro}-(a)), which integrate multimodal signals into a relational structure and have proven effective in knowledge-centric tasks such as KG completion~\cite{mose, imf, native, zhao2024contrast, mmkgsurvey}.
However, existing MMKGs are built for broad, open-domain sources such as Freebase or Wikidata and thus fail to capture the domain-specific semantics and modality patterns required in e-commerce.

Building on this motivation, we design \emmkgr\ with three key components as follows:

\begin{itemize}[leftmargin=*, topsep=0pt]
    \item \textbf{{\emmkg} Construction:}
    As illustrated in Figure~\ref{fig:intro}-(b), {\emmkg} contains \textit{item nodes} and \textit{modality nodes}, where each modality node corresponds to a specific modality instance of an item (\eg, image, description, or review).
    In an {\emmkg}, item-modal edges link each item to its own modality instances, while modal-modal edges capture semantic similarity between modality instances belonging to different items.
    This simple relational structure naturally supports modality extensibility (L1).

    \item \textbf{Unified Representation Learning:}
    We employ a Graph Neural Network (GNN)~\cite{lightgcn} that propagates multimodal and relational signals through \emmkg\ and incorporates a KG-oriented objective~\cite{rotate} to encode relation-level semantics, thereby enabling each \textit{item embedding} to aggregate information from its modality instances and semantically related neighbors.
    Consequently, the resulting representations naturally generalize across tasks (L2).
    
    \item \textbf{Downstream Applications:}
    The learned {\emmkg} representations can be readily applied to various e-commerce tasks. 
    We demonstrate their utility through two representative examples, \ie, recommendation and product search, and provide qualitative visualization to analyze the learned semantic structure.
\end{itemize}

\vspace{1mm}
\noindent We clarify that this work aims to establish and validate a new, data-driven design direction for addressing (L1) and (L2), emphasizing feasibility and conceptual insight \textit{rather than technical depth}.

\vspace{1mm}
\noindent\textbf{Contributions.}
Our contributions are summarized as follows:
\begin{itemize}[leftmargin=*]
    \item \textbf{New Direction for MMKG in E-commerce:}
    We introduce {\emmkg}, the first multimodal knowledge graph for e-commerce, offering a new perspective on modeling e-commerce items.
    \item \textbf{Unified Representation Framework:}
    We develop \emmkgr, a unified representation framework that simultaneously addresses modality extensibility (L1) and task generalization (L2) through a shared semantic space.
    \item \textbf{Empirical Validation and Analysis:}
    We demonstrate consistent gains of up to 10.18\% in Recall@10 for recommendation and up to 21.72\% over vector-based retrieval for product search on Amazon datasets, and provide qualitative visualization to reveal the semantic organization of the learned representations.
\end{itemize}

\begin{figure*}[t]
\centering
\includegraphics[width=\linewidth]{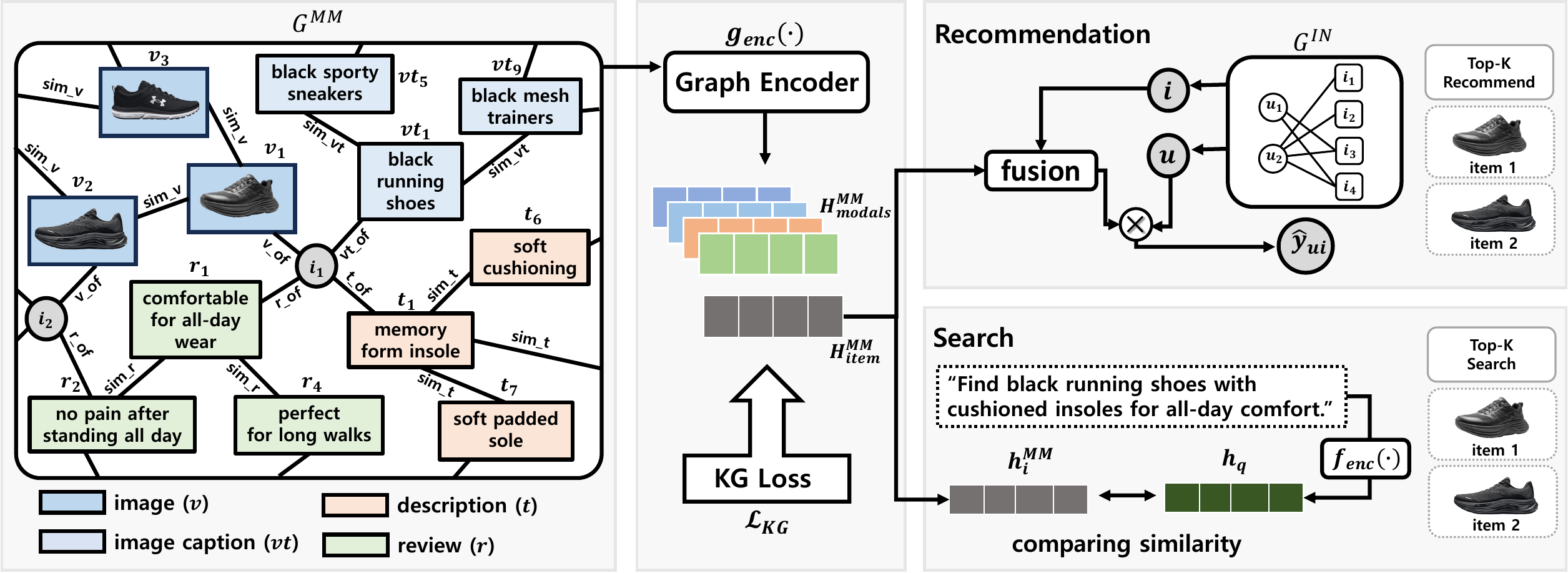}
\caption{Overview of the \emmkgr\ framework.}
\label{fig:overview}
\end{figure*}
\section{Related Work} \label{sec:rworks} \noindent\textbf{Multimodal Recommender Systems.}
MMRSs~\cite{grcn,dualgnn,lattice,bm3,freedom,lgmrec,mmrec_survey} enhance collaborative filtering by incorporating 
item-side multimodal signals.
A common approach constructs item–item graphs using modality-specific similarities~\cite{lattice,freedom}, later aggregated across modalities.
However, because each modality is modeled \textit{independently}, these methods cannot capture cross-modal relationships.

\vspace{1mm}
\noindent\textbf{Multimodal Knowledge Graphs.}
MMKGs augment knowledge graphs with multimodal signals, 
enabling richer relational reasoning~\cite{mmkgsurvey, mose, imf, native, zhao2024contrast}. 
While they have shown strong performance in multimodal KG completion~\cite{mose, imf, native, zhao2024contrast} and entity alignment~\cite{ent_align, chen2025noise, psnea, eiea}, prior work has focused mainly on \textit{general-purpose} knowledge domains, leaving e-commerce-specific MMKGs largely underexplored. 
 
\section{Methodology} \label{sec:method} In this section, we describe the construction of {\emmkg}, the unified representation learning process, and their use as a flexible backbone for downstream e-commerce tasks.

\subsection{{\emmkg} Construction} \label{sec:method-mmkg}

\vspace{1mm}
\noindent\textbf{Graph Definition.}
Let $\mathcal{I}={i_1, i_2, \ldots, i_N}$ denote the set of item nodes, where $N=|\mathcal{I}|$.
Each item $i_j$ is associated with multimodal content, for which we create \textit{modality-instance nodes}.
For each modality type $\tau \in \{v, t, r, vt\}$ representing \textit{image}, \textit{description}, \textit{review}, and \textit{image caption}, respectively, we define the set \mbox{$\mathcal{M}^{\tau}\!=\!\{m_1^{\tau},\ldots,m_N^{\tau}\}$}, where each $m_j^{\tau}$ corresponds to the ${\tau}$-type modality instance of item $i_j$.
We then define the node set as $\mathcal{V}=\mathcal{I}\cup\mathcal{M}$, where $\mathcal{M}=\mathcal{M}^v\cup\mathcal{M}^t\cup\mathcal{M}^r\cup\mathcal{M}^{vt}$ denotes the union of all modality-instance nodes.
Similarly, we define the edge set as $\mathcal{E}=\mathcal{E}_{im}\cup\mathcal{E}_{mm}$, where $\mathcal{E}_{im}$ and $\mathcal{E}_{mm}$ denote item-modal and modal-modal edges, respectively. 
Putting these components together, the \emmkg\ is formally defined as $\mathcal{G}^{\mathrm{MM}} = (\mathcal{V}, \mathcal{E})$.

\vspace{1mm}
\noindent\textbf{Initial Node Representation.}
Item node $i_j$ is initialized with a learnable ID embedding $\mathbf{e}_i$, and for each modality-instance node $m_j^{\tau}$, we extract modality features using a pretrained encoder $f_{\mathrm{enc}}(\cdot)$:

\vspace{-0.2cm}
\footnotesize
\begin{equation}
{\hat{m}}^{\tau}_{j} = f_{\mathrm{enc}}(m_j^{\tau}), \quad \tau \in \{v, t, r, vt\}.
\end{equation}
\normalsize
This design naturally supports modality extensibility by adding new modality-instance nodes with their corresponding encoders.

\vspace{1mm}
\noindent\textbf{Relational Structure.}
Edges in $\mathcal{G}^{\mathrm{MM}}$ are designed to explicitly encode two types of relationships central to multimodal item understanding:
(1) \textit{item–modal} edges $\mathcal{E}_{im}$, which link each item node $i_j$ to \textit{its} modality-instance nodes ($m_j^v$, $m_j^t$, $m_j^r$, $m_j^{vt}$), capturing the factual relationship between an item and its multimodal attributes (\eg, \textit{image\_of});
(2) \textit{modal–modal} edges $\mathcal{E}_{mm}$, which connect modality-instance nodes belonging to \textit{different} items based on semantic similarity, allowing the graph to capture fine-grained multimodal relationships across the catalog (\eg, \textit{similar\_image}).
 To construct $\mathcal{E}_{mm}$, for each modality node $m_j^{\tau}$, we identify its top-$n$ most similar nodes $m_k^\tau$ according to cosine similarity:

\vspace{-0.2cm}
\footnotesize
\begin{equation}
(m_j^{\tau},m_k^{\tau})\in\mathcal{E}_{mm}\ \quad \text{if}\quad m_k^{\tau}\in
\mathrm{TopN}_{m_l^\tau\in\mathcal{M}^\tau \setminus \{m_j^\tau\}} 
\big(\mathrm{sim}(\hat{m}_j^{\tau},\hat{m}_l^{\tau})\big).
\end{equation}
\normalsize
The resulting adjacency matrix of $\mathcal{G}^{\mathrm{MM}}$ can be written as follows:
\footnotesize

\begin{equation}
\textbf{A}^{\mathrm{MM}}=
\begin{bmatrix}
0 & \mathbf{A}_{\mathrm{im}} \\
\mathbf{A}_{\mathrm{im}}^{\top} & \mathbf{A}_{\mathrm{mm}}
\end{bmatrix},
\end{equation}
\normalsize
where $\mathbf{A}_{\mathrm{im}}\in\mathbb{R}^{|\mathcal{I}|\times|\mathcal{M}|}$ and 
$\mathbf{A}_{\mathrm{mm}}\in\mathbb{R}^{|\mathcal{M}|\times|\mathcal{M}|}$ represent item–modal connections and modal similarity links, respectively.
This design unifies item attributes and cross-item multimodal relationships in a single graph and is  validated through ablation against alternative structural variants (\eg, adding \textit{item-item} edges $\mathbf{A}_{\mathrm{ii}}$) in Section~\ref{sec:exp_results}

\subsection{Unified Representation Learning} \label{sec:method-mmkgr}
This module learns unified node representations that reflect multimodal neighborhood patterns and relation-level semantics.

\vspace{1mm}
\noindent\textbf{Graph Encoder.}
To propagate multimodal and relational signals across $\mathcal{G}^{\mathrm{MM}}$, we use a LightGCN-based encoder $g_{\mathrm{enc}}(\cdot)$ due to its simplicity and strong empirical performance~\cite{lightgcn}.
To this end, we project $\hat{m}_j^\tau$ into a shared embedding space via $z_j^\tau = f_{\mathrm{proj}}(\hat{m}_j^\tau)$, $\tau \in \{v,t,r,vt\}$, and construct the full embedding matrix as 
$\mathbf{E}^{MM} = [\mathbf{E}^i; \mathbf{Z}^v; \mathbf{Z}^t; \mathbf{Z}^r; \mathbf{Z}^{vt}] \in \mathbb{R}^{5N \times d}$, where $\mathbf{Z}^\tau$ stacks $\{z_j^\tau\}_{j=1}^{N}$, $\mathbf{E}^i$ denotes the learnable ID embeddings of item nodes, $N = |\mathcal{I}|$, and $d$ is the embedding dimension. 
Over $L$ layers, the encoder updates node embeddings through multiplication with the normalized adjacency matrix and aggregation via averaging across layers:

\vspace{-0.1cm}
\footnotesize
\begin{equation}
\mathbf{H}^{MM} = \frac{1}{L+1}\sum_{l=0}^{L}\mathbf{H}^{(l)},\qquad \mathbf{H}^{(l)}=\widetilde{\mathbf{A}}^{MM}\mathbf{H}^{(l-1)}, \quad l > 0, \quad \mathbf{H}^{(0)}=\mathbf{E}^{MM},
\end{equation}
\normalsize
where $\widetilde{\mathbf{A}}^{MM}$ denotes the symmetric normalized adjacency matrix. 
This formulation is lightweight and model-agnostic, allowing stronger encoders to be integrated without altering the graph design.

\vspace{1mm}
\noindent\textbf{Knowledge Graph Loss.}
To capture relation-level semantics, we incorporate a KG-oriented objective based on RotatE~\cite{rotate}, which models relations as rotations in a complex space.
Given a triple $(h, r, t)$, the scoring function is defined as $f(h, r, t) = \|\mathbf{h} \circ \mathbf{r} - \mathbf{t}\|^{2},$
where $\mathbf{h}$ and $\mathbf{t}$ denote the embeddings of the head and tail entities (\ie, item or modality nodes). 
Each relation type (\eg, $image\_of$, $similar\_image$) is associated with a learnable embedding $\mathbf{r}$, which is randomly initialized and optimized during training.
The loss $\mathcal{L}_{\mathrm{KG}}$ is defined as follows~\cite{rotate}:

\footnotesize
\begin{equation}
\mathcal{L}_{\mathrm{KG}}
= \sum_{(h, r, t, t') \in \mathcal{T}}
\log \sigma\!\big(f(h,r,t) - f(h,r,t')\big),
\end{equation}
\normalsize
where $\mathcal{T}$ denotes positive triples paired with their negative counterparts and $\sigma(\cdot)$ is the activation function.
Jointly optimizing this loss with the graph encoder yields unified multimodal representations $\mathbf{H}^{\mathrm{MM}}$ that capture both neighborhood structure and relation semantics, providing a shared semantic basis for downstream tasks.

\subsection{Downstream Applications} \label{sec:method-app}
We illustrate how the unified representations learned from {\emmkg} can be applied to representative e-commerce tasks.
These examples show that once a multimodal semantic space is constructed, the multimodal item embeddings can be easily incorporated into existing task models without requiring major architectural changes.

\vspace{1mm}
\noindent\textbf{Recommendation.}
We enhance collaborative filtering (CF) by augmenting it with the multimodal item embeddings $\mathbf{H}_{\mathrm{item}}^{\mathrm{MM}}$ obtained from {\emmkg}.
A user–item interaction graph $\mathcal{G}^{\mathrm{IN}}$ is encoded using the same LightGCN-based encoder $g_{\mathrm{enc}}(\cdot)$ to obtain user embeddings $\mathbf{H}_{\mathrm{user}}^{\mathrm{IN}}$ and interaction-based item embeddings $\mathbf{H}_{\mathrm{item}}^{\mathrm{IN}}$.
The interaction-based item embeddings are combined with multimodal item embeddings to form unified item representations, \ie, $\mathbf{H}_{\mathrm{item}} = \mathbf{H}_{\mathrm{item}}^{\mathrm{IN}} + \mathbf{H}_{\mathrm{item}}^{\mathrm{MM}}$.
Given a user–item pair $(u,i)$, the predicted score is computed as $\hat{y}_{ui} = \mathbf{h}_u^{\mathrm{IN}} \cdot \mathbf{h}_i$.
The model is optimized using the Bayesian Personalized Ranking (BPR) loss~\cite{rendle2009bpr}: 

\footnotesize
\[
\mathcal{L}_{\mathrm{BPR}} = 
-\!\!\sum_{(u,i,j)\in\mathcal{R}}
\log \sigma \big(\hat{y}_{ui} - \hat{y}_{uj}\big),
\tag{6}
\]
\normalsize
where $\mathcal{R}$ contains training triples of user $u$, positive item $i$, and negative item $j$.

\vspace{1mm}
\noindent\textbf{Search.} 
For product retrieval, a query $q$ is encoded using the same pretrained modality encoder $f_{\mathrm{enc}}(\cdot)$ (\eg, text or image) introduced in Section~\ref{sec:method-mmkg}, yielding a query embedding $\mathbf{h}_q$.
Retrieval is then performed by ranking items according to cosine similarity with the multimodal item embeddings $\mathbf{h}_i^{MM}$:

\vspace{-0.1cm}
\footnotesize
\[
\mathcal{I}_{\mathrm{top}\text{-}N} 
= \mathrm{TopN}_{i\in\mathcal{I}}\big(
\mathrm{sim}(\mathbf{h}_q, \mathbf{h}_i^{MM})\big).
\tag{7}
\]
\normalsize
where $\mathrm{sim}(\cdot,\cdot)$ denotes the cosine similarity function.

\begin{table}[t]
\centering
\footnotesize
\caption{Dataset statistics}
\label{tab:dataset-stats}
\resizebox{\linewidth}{!}{%
\begin{tabular}{lcccccc}
\toprule
\textbf{Statistic} & \textbf{Office} & \textbf{Grocery} & \textbf{Pet} & \textbf{Toys} & \textbf{Beauty} & \textbf{Clothing} \\
\midrule
\textbf{\# User} & 4{,}905 & 14{,}681 & 19{,}856 & 19{,}412 & 22{,}363 & 39{,}387 \\
\textbf{\# Item} & 2{,}420 & 8{,}713 & 8{,}510 & 11{,}924 & 12{,}101 & 23{,}033 \\
\textbf{\# Interaction} & 53{,}258 & 151{,}254 & 157{,}836 & 167{,}597 & 198{,}502 & 278{,}677 \\
\bottomrule
\end{tabular}}
\end{table}
\section{Evaluation} \label{sec:eval} \subsection{Experimental Setup}\label{sec:exp_setup}

\vspace{1mm}
\noindent\textbf{Datasets.}
Experiments are conducted on six Amazon subsets\footnote{\url{https://cseweb.ucsd.edu/~jmcauley/datasets/amazon/links.html}.}--Office, Grocery, Pet, Toys, Beauty, and Clothing (see Table~\ref{tab:dataset-stats})--which include user–item interactions and multimodal content.
Image features are extracted using a pretrained \textit{CNN}~\cite{deepcnn}, and textual features using \textit{SBERT}~\cite{sbert}.
To better reflect diverse real-world product information, we additionally include two LLM-derived modalities: (1) review summaries generated via \textit{gpt-4o-mini}, and (2) image captions from the Amazon Review Plus Dataset~\cite{imgc_dataset}.
Both modalities are encoded using the \textit{OpenAI text-embedding-3-large} model.

\vspace{1mm}
\noindent\textbf{Baselines.}
For \textbf{recommendation}, we compare {\emmkgr} with LightGCN~\cite{lightgcn} and six MMRS models: GRCN~\cite{grcn}, DualGNN~\cite{dualgnn}, LATTICE~\cite{lattice}, BM3~\cite{bm3}, FREEDOM~\cite{freedom}, and LGMRec~\cite{lgmrec}.
For \textbf{product search}, we use a simple vector-based retrieval baseline~\cite{mmRAG_survey} that compares a query embedding to each modality embedding independently and ranks items based on the best single-modality match. 
This baseline does not integrate multimodal signals.

\vspace{1mm}
\noindent\textbf{Evaluation Protocol.}
For \textbf{recommendation}, we follow an 80/10/10 train–validation–test splitting and report Recall@$K$ and NDCG@$K$.
For \textbf{search}, we construct two query sets: 
(1) \textbf{fine-grained} queries describing modality-specific attributes of a single target item (\eg, \textit{``Find black running shoes with cushioned insoles for all-day comfort''}); (2) \textbf{coarse-grained} queries expressing broader intent shared across an item cluster (\eg, \textit{``Find comfortable black running shoes''}). Search performance is evaluated with Recall@$K$ and MAP@$K$.

\begin{table}[t]
\centering
\footnotesize
\caption{Ablation results of the proposed {\emmkg} on recommendation (R@10) and search (MAP@10) for Clothing}
\label{tab:eq1_r10_map10}
\resizebox{0.75\linewidth}{!}{%
\begin{tabular}{lccc}
\toprule
\multirow{2}{*}{\textbf{Variant}} &
\multirow{2}{*}{\textbf{Recommendation}} &
\multicolumn{2}{c}{\textbf{Search}} \\
\cmidrule(lr){3-4}
 &  & Fine & Coarse \\
\midrule
original                & \textbf{0.0611} & \textbf{0.5813} & \textbf{0.6839} \\
\textit{w/ interaction} & 0.0595          & 0.5086          &  0.6296         \\
\textit{w/ inter-modal} & 0.0601          & 0.5643          &  0.6805       \\
\textit{w/ item-item}   & 0.0580         &  0.5386         &  0.6538        \\
\bottomrule
\end{tabular}%
}
\end{table}
\begin{table}[t]
\centering
\footnotesize
\caption{Recommendation comparison (R@10)}
\label{tab:eq2_r10}
\resizebox{.98\linewidth}{!}{%
\begin{tabular}{lcccccc}
\toprule
\textbf{Method} & \textbf{Office} & \textbf{Grocery} & \textbf{Pet} & \textbf{Toys} & \textbf{Beauty} & \textbf{Clothing} \\
\midrule
LightGCN  & 0.0767 & 0.0922 & 0.0839 & 0.0845 & 0.0832 & 0.0361 \\
GRCN      & 0.0856 & 0.1008 & 0.0976 & 0.0874 & 0.0918 & 0.0428 \\
DualGNN   & 0.0850 & 0.1065 & 0.0916 & 0.0867 & 0.0917 & 0.0438 \\
LATTICE  & 0.0920 & 0.1056 & 0.1042 & 0.0940 & 0.0940 & 0.0493 \\
BM3       & 0.0706 & 0.0876 & 0.0907 & 0.0761 & 0.0865 & 0.0427 \\
FREEDOM  & 0.0930 & 0.1197 & \underline{0.1157} & \underline{0.1080} & \underline{0.1079} & \textbf{0.0621} \\
LGMRec   & \underline{0.0963} & \underline{0.1201} & 0.1106 & 0.1005 & 0.1072 & 0.0538 \\
\textbf{\emmkgr} & \textbf{0.1061} & \textbf{0.1235} & \textbf{0.1187} & \textbf{0.1101} & \textbf{0.1133} & \underline{0.0611} \\
\bottomrule
\end{tabular}}
\end{table}

\vspace{-0.1cm}
\subsection{Experimental Results}\label{sec:exp_results}

\noindent\textbf{Effectiveness of the {\emmkg} Structure.}
Ablation experiments compare our {\emmkg} with three variants:
(1) \textit{w/ interaction}, which injects user–item interaction edges directly into the MMKG;
(2) \textit{w/ inter-modal}, which directly connects all modality nodes associated with the same item to each other (\eg, its image, description, and review); and
(3) \textit{w/ item-item}, which uses direct item–item edges constructed from modality-specific similarities with late aggregation~\cite{lattice, freedom}.
The results in Table~\ref{tab:eq1_r10_map10} show that our design consistently yields the best performance across both recommendation and search. 
These results can be explained by three structural limitations:
(1) adding interaction edges injects behavioral noise into multimodal propagation, (2) inter-modality edges lead to over-aggregation that collapses modality distinctions, and (3) direct item-item edges miss fine-grained multimodal relationships by bypassing modality nodes that encode modality-specific similarities.

\vspace{1mm}
\noindent\textbf{Recommendation Performance.}
Across all datasets 
(Table~\ref{tab:eq2_r10}), {\emmkgr} achieves state-of-the-art recommendation performance, improving Recall@10 by up to 10.18\%. 
Notably, {\emmkgr} achieves either the best or second-best performance across all datasets, demonstrating robust effectiveness over diverse product domains.
These results indicate that the unified multimodal representations complement CF signals more effectively than existing MMRS models, especially those relying on independently constructed modality graphs (\eg, LATTICE and FREEDOM).

\vspace{1mm}
\noindent\textbf{Generalization to Product Search.}
In product search, {\emmkgr} substantially outperforms vector-based retrieval, achieving gains of 7.80\% on fine-grained queries and 21.72\% on coarse-grained queries.
These improvements show that \emmkg\ embeds products in a unified multimodal semantic space that supports both fine-grained attribute matching and broader semantic retrieval, whereas vector-based methods struggle to model such relationships when treating modalities independently.

\begin{table}[t]
\centering
\footnotesize
\caption{Search comparison for Clothing
}
\label{tab:eq3}
\resizebox{0.9\linewidth}{!}{%
\begin{tabular}{lcccc}
\toprule
\multirow{2}{*}{Metric} & \multicolumn{2}{c}{\textbf{Fine-grained Query}} & \multicolumn{2}{c}{\textbf{Coarse-grained Query}} \\
\cmidrule(lr){2-3} \cmidrule(lr){4-5}
 & \textbf{Vector-based} & \textbf{E-MMKGR} & \textbf{Vector-based} & \textbf{E-MMKGR} \\
\midrule
R@5  & 0.6538 & \textbf{0.7048} & 0.6297 & \textbf{0.7665} \\
R@10 & 0.7514 & \textbf{0.7789} & 0.7905 & \textbf{0.8759} \\
MAP@5  & 0.4966 & \textbf{0.5713} & 0.4923 & \textbf{0.6423} \\
MAP@10 & 0.5096 & \textbf{0.5813} & 0.5448 & \textbf{0.6839} \\
\bottomrule
\end{tabular}%
}
\end{table}

\begin{figure}[t]
\centering
\includegraphics[width=\linewidth]{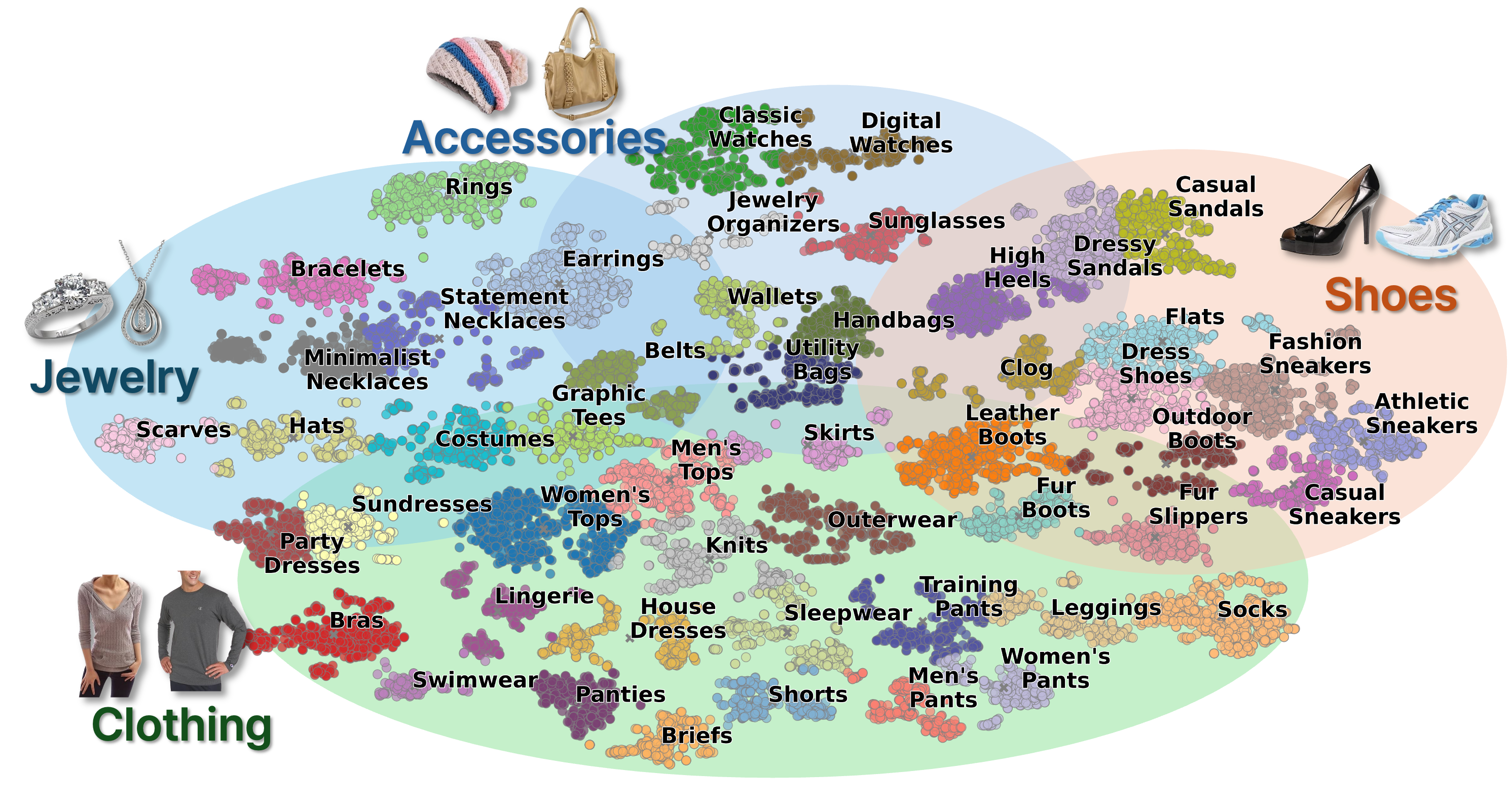}
\caption{t-SNE visualization of unified item representations for Clothing with K-means clustering}
\label{fig:vis_cluster}
\end{figure}

\vspace{1mm}
\noindent\textbf{Semantic Cohesion of Unified Representations.}
We qualitatively analyze whether the learned unified representations capture multimodal product semantics using t-SNE and $K$-means clustering ($K=50$, determined via the elbow method~\cite{elbow}).
As shown in Figure~\ref{fig:vis_cluster}, items are primarily grouped by high-level categories (\eg, clothing, shoes, jewelry, and accessories), while each cluster exhibits semantically continuous structures that reflect coherent alignment across diverse modalities.
For instance, within the \textit{Shoes} category, clusters form a smooth spectrum ranging from formal footwear (\eg, high heels) to casual and sports-oriented footwear (\eg, sneakers and running shoes), suggesting that 
structured multimodal integration preserves fine-grained stylistic and functional semantics in the unified representation.

We further assess cluster-level semantic cohesion by comparing intra- and inter-cluster cosine similarities based on item embeddings.
We report these measures for both unified multimodal embeddings and unimodal embeddings derived from individual modalities (\eg, image features $\hat{m}_j^v$).
Notably, the unified multimodal representation yields the largest intra-inter similarity gap, reflecting stronger cluster separation and fine-grained product semantics.
\section{Conclusion} \label{sec:concl} We introduced {\emmkg}, the first multimodal knowledge graph for e-commerce, and {\emmkgr}, a unified framework built upon it.
By explicitly modeling item–modality and cross-item relationships, {\emmkgr} learns extensible multimodal representations that support diverse downstream tasks while preserving fine-grained product semantics.
Comprehensive experiments on recommendation and product search, together with qualitative analyses of semantic cohesion and cluster structure, demonstrate its effectiveness and robust generalization across diverse product domains.

\balance
\bibliographystyle{ACM-Reference-Format}
\bibliography{bibliography}

\clearpage
\setcounter{table}{0}
\setcounter{figure}{0}

\end{document}